# Initial stage of cavitation in liquids and its observation by Rayleigh scattering

**M. Pekker[1] and M. N. Shneider[2,*]**

[1] *Department of Mechanical and Aerospace Engineering,* George Washington University, *Washington, DC 20052*
[2] *Department of Mechanical and Aerospace Engineering, Princeton University, Princeton, NJ, 08544*

A theory is developed for the initial stage of cavitation in the framework of Zel'dovich-Fisher theory of nucleation in the field of negative pressure, while taking into account the surface tension dependence on the nanopore radius. A saturation mechanism is proposed that limits the exponential dependence of the nucleation rate on the energy required to create nanopores. An estimate of the saturated density of nanopores at the nucleation stage is obtained. It is shown that Rayleigh scattering can detect nanopores arising at the initial stage of cavitation development.

## 1. Introduction

The question regarding cavitation bubbles occurrence in negative pressure regions has had a century-long history, but it still have not had a clear enough answer and explanation until now. Indeed, based on the theory of Zel'dovich-Fisher [1,2], the probability of the creation of the critical bubble due to the presence of thermal fluctuations has the form (see also [3-6]):

$$W_{pore} = 1 - \exp\left( -\int_0^t \int_V \Gamma dt_1 dV \right), \tag{1}$$

where $V$ and $t$ are the volume and the duration of the measurement, respectively. $\Gamma\,[\mathrm{m^{-3}s^{-1}}]$ in (1) characterizes the rate of creation of the cavitation voids per unit volume per second:

$$\frac{dn_{cr}}{dt} = \Gamma = \frac{1}{V_{cr}}\frac{1}{\tau_{\exp}} = \Gamma_0 \exp\left( -\frac{W_{cr}}{k_B T} \right), \quad [\mathrm{m^{-3}s^{-1}}] \tag{2}$$

where $n_{cr}$ is the density of bubbles with critical radius $R_{cr}$, $V_{cr} = 4\pi R_{cr}^3/3$, $\tau_{\exp}$ is the expectation time for appearance of the pore with the critical radius, $k_B$ is the Boltzmann constant, and $T$ is the temperature of the fluid, $W_{cr}$ is the fluctuation of energy corresponds to the fluctuation of the radius $R_{cr}$, $\Gamma_0$ is the kinetic prefactor, which depends on the theoretical model used in [1-6]. For example, in [3, 6], it was accepted that

$$\Gamma_0 \approx \frac{1}{V_{cr}}\frac{1}{\Delta t_T} = \frac{3}{4\pi R_{cr}^3}\frac{k_b T}{2\pi\hbar}, \tag{3}$$

where $\frac{1}{\Delta t_T} = \frac{k_B T}{2\pi\hbar}$ is the effective thermal frequency; $\hbar$ is the Planck constant. According to the Zel'dovich-Fischer theory:

$$W_{cr} = \frac{16\pi\sigma_s^3}{3P_-^2}, \tag{4}$$

* m.n.shneider@gmail.com

where $\sigma_s$ is the surface tension coefficient, and $P_-$ is the negative pressure in liquid. The critical radius is $R_{cr}$, and it is connected with $\sigma_s$ and $P_-$ by the relationship:

$$R_{cr} = \frac{2\sigma_s}{|P_-|} . \tag{5}$$

Substituting the expressions (3) and (4) into (2), we obtain:

$$\frac{dn_b}{dt} = \Gamma = \frac{3|P_-|^3}{16\pi\sigma_s^3}\frac{k_B T}{2\pi\hbar}\exp\left(-\frac{16\pi\sigma_s^3}{3k_B T P_-^2}\right) \tag{6}$$

The critical pressure, at which cavitation occurs can be easily estimated by equating the exponential factor in (6) to one:

$$|P_{cr}| \sim \sqrt{16\pi\sigma_s^3 / 3k_B T} . \tag{7}$$

Taking into account that $\sigma_s = 0.072\ \mathrm{N/m^2}$ for water at the temperature $T \approx 300\,\mathrm{K}$ [7], the corresponding $P_{cr} \approx -1200\,\mathrm{MPa}$. Accounting for pre-exponential factor in (6) has practically no effect for the estimation (7). The critical pore radius corresponding to the value $P_{cr} \approx -1200\,\mathrm{MPa}$ is $R_{cr} \approx 1.2\cdot 10^{-10}\,\mathrm{m}$, that is of order of the averaged radius of water molecules.

The experimentally measured values of the negative pressure, at which the cavitation bubbles occur in water, are in a wide range from -2MPa up -140MPa [8-15]. At present, the conventional negative pressure value, at which the water starts cavitation, is $P_{cr} \approx -30\,\mathrm{MPa}$.

## 2. Zel'dovich-Fisher theory modification

Naturally, the question arises: what is the reason for such a strong discrepancy between theory and experiment? Equation (6) was obtained under the assumption of constant surface tension $\sigma_s$, which corresponds to an infinitely thin boundary that separates the vacuum pore (or vapor bubble) from the liquid. However, as was first noted by Tolman [16], an approximation of the infinitely thin boundary is valid only when the thickness of the transition layer $\delta_b$ is much smaller than the pore radius $R_b$. Evidently, when the pore radius $R_b$ is of the order of $\delta_b$, the surface tension $\sigma_s$ should decrease and tend to zero at $R_b \to 0$. Tolman proposed the following formula for the surface tension coefficient

$$\sigma_s = \frac{\sigma_{0s}}{1+\delta_b / R_b} , \tag{8}$$

Since at $\delta_b = 0$, the surface energy of the pore is $W_b = 4\pi\sigma_{0s}R_b^2$, the correction to that is related to the size of the transition layer $\delta_b$ should be about $\delta W_b = -4\pi\sigma_{0s}\delta_b^2$. Based on this, we suggest expressing the approximation formula for the surface tension coefficient $\sigma_s$ in "Lorentzian" form:

$$\sigma_s = \frac{\sigma_{0s}}{1+\delta_b^2 / R_b^2} . \tag{9}$$

Fig.1 shows the dependence $\sigma_S/\sigma_{0s}$ on $\delta_b/R_b$ for Tolman's (8) and the "Lorentzian" (9) approximations.

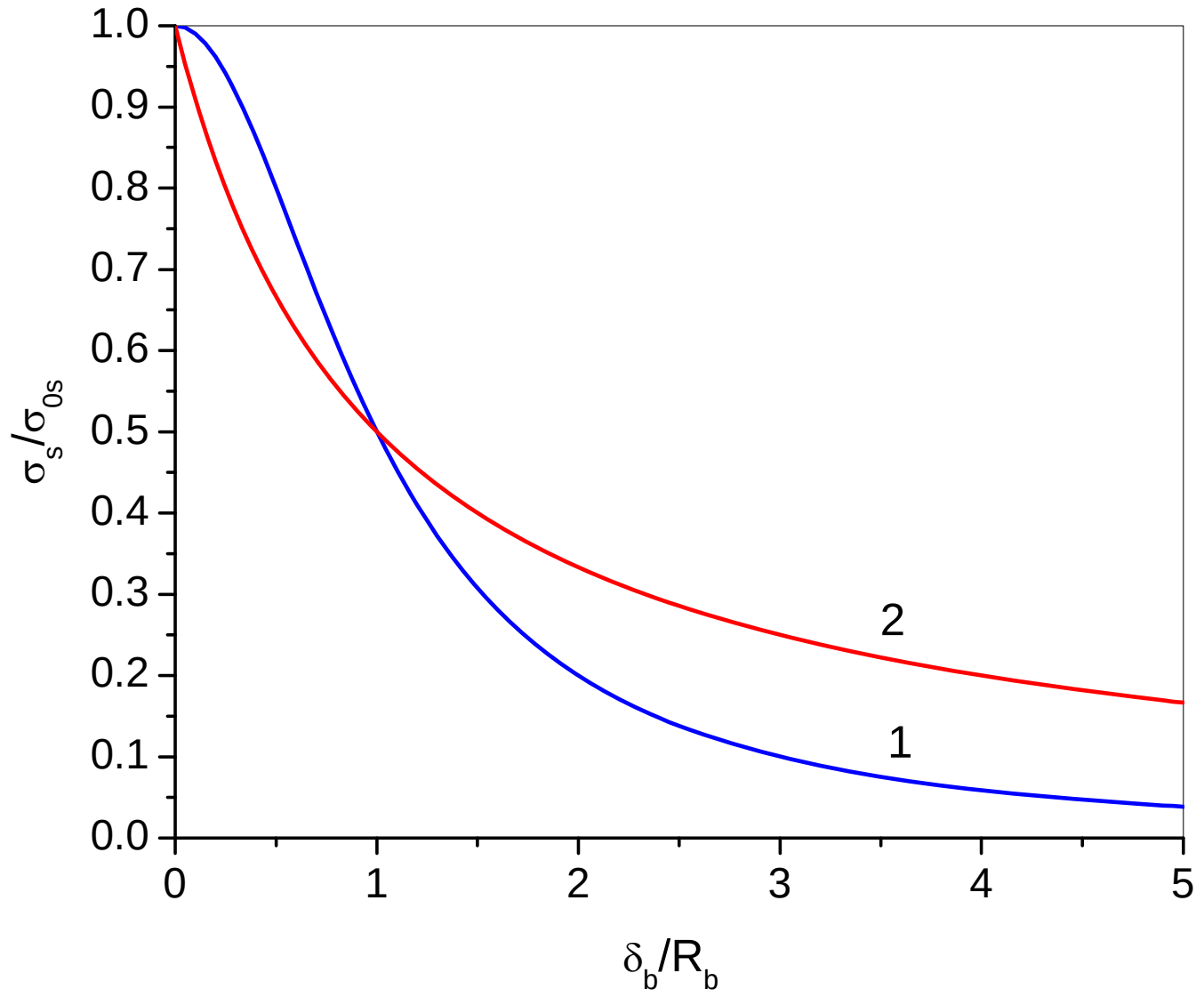

**Fig. 1** Dependence of $\sigma_s/\sigma_{0s}$ on $\delta_b/R_b$ in the "Lorentzian" approximation (2.10) (curve 1) and the Tolman approximation (2.9) (curve 2).

Further, all the calculations will be carried out with the surface tension, taken in the assumption (9). In accordance with [1] and [2], the energy required to create a bubble of the radius $R_b$ in a liquid is:

$$W(R_b) = -\int_0^{R_b} 4\pi r^2 |P_-| dr + \int_0^{R_b} 8\pi r^2 \frac{\sigma_s}{r} dr \,. \quad (10)$$

Substituting the value $R_b$ from (10), we obtain

$$W(R_b) = -\frac{4\pi}{3}|P_-|R_b^3 + 4\pi R_b^2 \sigma_{0s} - 4\pi\sigma_{0s}\delta_b^2 \ln\left(1+\frac{R_b^2}{\delta_b^2}\right). \quad (11)$$

From (11), it is easy to find the value of the radius $R_{cr}$ at which the energy $W(R_b)$ reaches its maximum at a fixed value of $P_-$:

$$R_{cr} = \frac{\sigma_{0s}}{|P_-|} + \sqrt{\left(\frac{\sigma_{0s}}{P_-}\right)^2 - \delta_b^2} = \frac{\sigma_{0s}}{|P_-|}\left(1+\sqrt{1-\frac{\delta_b^2 P_-^2}{\sigma_{0s}^2}}\right) = \frac{R_{cr,0}}{2}\left(1+\sqrt{1-\frac{4\delta_b^2}{R_{cr,0}^2}}\right), \; R_{cr,0} = \frac{2\sigma_{0s}}{|P_-|} \quad (12)$$

From (13) it follows that $R_{cr}$ depending on the $\delta_b$, is within the range:

$$\frac{1}{2} \le \frac{R_{cr}}{R_{cr,0}} = \frac{1}{2}\left(1+\sqrt{1-\frac{4\delta_b^2}{R_{cr,0}^2}}\right) < 1 \quad (13)$$

The corresponding critical energy at this radius is

$$W_{cr} = -\frac{4\pi}{3}|P_-|R_{cr}^3 + 4\pi R_{cr}^2 \sigma_{0s}\left(1-\frac{\delta_b^2}{R_{cr}^2}\ln\left(1+\frac{R_{cr}^2}{\delta_b^2}\right)\right) \,. \quad (14)$$

Equations (14) can be rewritten as

$$\widetilde{W}_{cr} = \frac{W_{cr}}{2\pi \cdot |P_{-}| R_{cr,0}^{3}/3} = \left( 3\xi^{2}\left(1 - \frac{\delta_b^{2}}{R_{cr,0}^{2}\xi^{2}} \ln\left(1 + \frac{R_{cr,0}^{2}\xi^{2}}{\delta_b^{2}}\right)\right) - 2\xi^{3} \right), \quad \xi = \frac{R_{cr}}{R_{cr,0}} = \frac{R_{cr}|P_{-}|}{2\sigma_0} \tag{15}$$

Dependences of the parameters $\xi = R_{cr}/R_{cr,0}$ and $\widetilde{W}_{cr}$ on $\delta_b / R_{cr,0}$ are shown in Fig. 2.

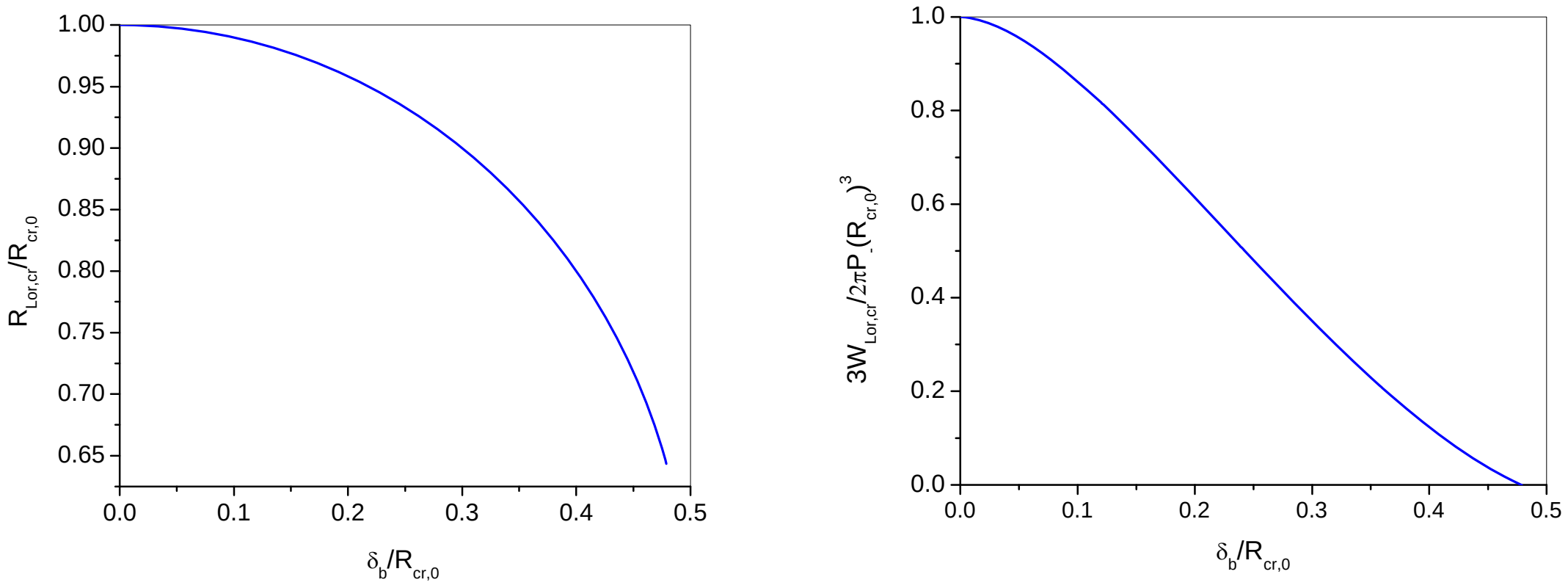

**Fig. 2** Dependences of the parameters $R_{cr}/R_{cr,0}$ and $\dfrac{W_{cr}}{2\pi \cdot |P_{-}| \cdot R_{cr,0}^{3}/3}$ on $\delta_b / R_{cr,0}$ .

It is easy to see that at $|P_{-}| << \sigma_{0s}/\delta_b$ ,

$$R_{cr} \approx \frac{2\sigma_{0s}}{|P_{-}|} \quad \text{and} \quad W_{cr} \approx \frac{2\pi R_{cr,0}^{3}|P_{-}|}{3} = \frac{16\pi}{3}\frac{\sigma_{0s}^{3}}{P_{-}^{2}}, \tag{16}$$

which coincide with the values $W_{cr}$ and $R_{cr}$ as described by (4) and (5) respectively. Also, the formula for the number of the embryonic voids of critical radius produced per unit time per unit volume coincides with (6). The value $W_{cr} = 0$ at $\delta_b = 0.4785 R_{cr,0}$ corresponds to the barrier-free cavitation that is not dependent on the fluid temperature (Fig. 2).

Issues related to barrier-free cavitation were studied intensively for liquid helium. For example, it was shown in [17] that the critical negative pressure, at which cavitation develops in helium-3 and helium-4, is practically constant in temperature range 0.05-1K. A theoretical justification for the fact of the critical negative pressure independence on the temperature in liquid helium was given in [18].

Substituting (14) and (12) into (2) yields the equation for the growth rate of the number of pores per unit volume at the assumed surface tension coefficient (9):

$$\frac{dn_{cr}}{dt} = \Gamma = \frac{3}{4\pi R_{cr}^{3}} \frac{k_b T}{2\pi\hbar} \exp\left(-\frac{W_{cr}}{k_B T}\right), \quad [\mathrm{m^{-3}s^{-1}}] \tag{17}$$

Figs. 3 and 4 show dependencies of $R_{cr}$ and $\Gamma$ on $\delta_b$

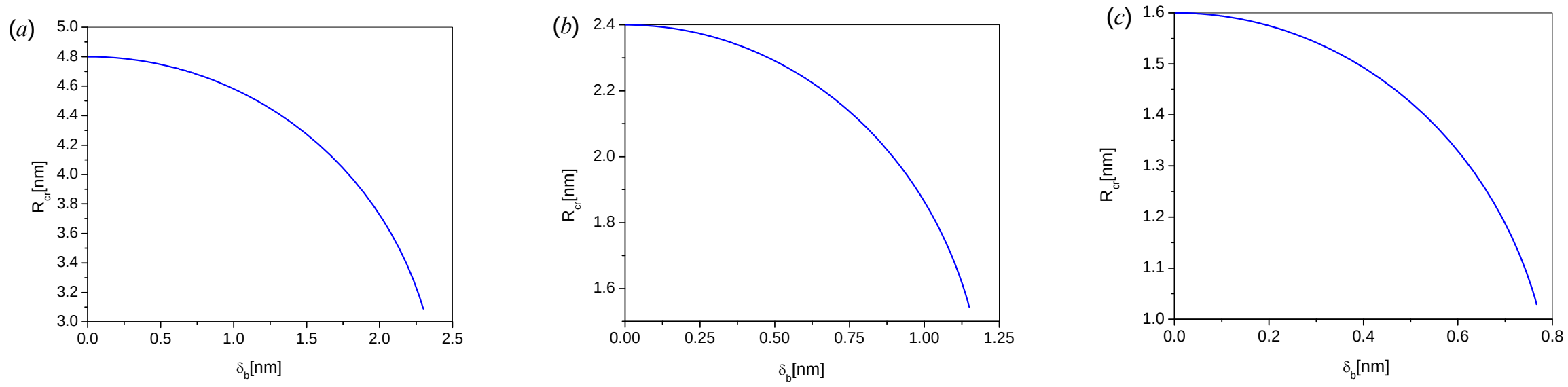

**Fig. 3**. Dependencies of $R_{cr}$ on $\delta_b$ at $P_{-} = -30\text{MPa}$ (a), $P_{-} = -60\text{MPa}$ (b), and $P_{-} = -90\text{MPa}$ (c).

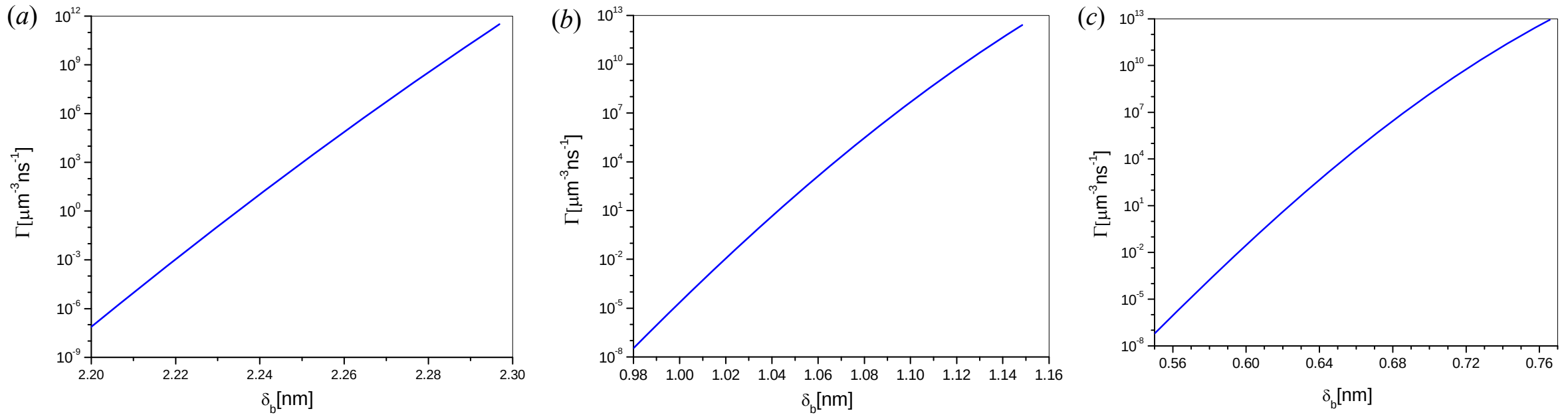

**Fig.4**. Dependencies of $\Gamma$ on $\delta_b$ at the same negative pressures as in Fig. 2.9.

From Fig.4 it follows that the amount of nanopores produced within one $\mu m^3$ can reach $10^{11} - 10^{13}$, that is an order of magnitude or more greater than the number of water molecules in this volume ( $n_{water} \approx 3.3 \cdot 10^{10} \mu m^{-3}$ ).

It is clear that this cannot be and, therefore, there must be a mechanism limiting the generation of nanopores.

### 3. A possible mechanism limiting the generation of cavitation nanopores

Let us consider a possible mechanism that limits the generation of cavitation nanopores. We will assume that the volume, in which a super-critical negative pressure occurs, is of the order of $l_{-}^3$. The formation of nanopores is associated with an increased volume of the cavitation region, which in turn leads to excessive positive pressure that reduces the value of the negative pressure in the area.

The number of nanopores that appeared in the negative pressure region $l_{-}^3$ over characteristic time of pressure equilibration $\tau = l_{-} / c_s$ ( $c_s$ is sound velocity) is equal to:

$$N \approx l_{-}^3 \int_0^{\tau} \Gamma dt \quad . \qquad (18)$$

Accordingly, the relative change in volume of liquid in the area where nanopores are emerging:

$$\frac{\delta V}{V} \approx \frac{V_{cr} N}{l_{-}^3} = V_{cr} n_b \quad , \qquad (19)$$

where $V_{cr}=\frac{4}{3}\pi R_{cr}^3$ is the volume of a nanpore of the critical radius $R_{cr}$ ($R_{cr}$ corresponds to the initial negative pressure $P_{-,0}$), and $n_b$ is density of this nanopores.

The value of the excess pressure $\delta P$ can be estimated from the simplest equation of state for a compressible fluid related to the sound velocity $c_s$:

$$\delta P = c_s^2 \delta\rho = c_s^2 \rho_0 \frac{\delta V}{V} = V_{cr} n_b c_s^2 \rho_0 , \tag{20}$$

where $\rho_0$ is the unperturbed density of the fluid.

The absolute value of the total pressure in the bubble generation region is equal to

$$\left|P_-\right| = \left|P_{-,0}\right| - \delta P = \left|P_{-,0}\right| - V_{cr} n_b c_s^2 \rho_0 . \tag{21}$$

Substituting $\left|P_-\right|$ into (15), we get

$$W_{cr} = W_{cr,0} + V_{cr}^2 c_s^2 \rho_0 n_b , \tag{22}$$

where $W_{cr,0}$

$$W_{cr,0} = -\frac{4\pi}{3}\left|P_-\right| R_{cr}^3 + 4\pi R_{cr}^2 \sigma_{0s}\left(1 - \frac{\delta_b^2}{R_{cr}^2}\ln\left(1 + \frac{R_{cr}^2}{\delta_b^2}\right)\right). \tag{23}$$

From (22), one can see that the increase in the number of pores per unit volume increases $W_{cr}$, and thus, reduces the rate of pore formation in (2).

Substituting (23) into (22) and, then into (17) yields:

$$\frac{dn_b}{dt} = \frac{3}{4\pi R_{cr}^3}\frac{k_B T}{2\pi\hbar}\exp\left(-\frac{W_{cr,0}}{k_B T}\right)\cdot\exp\left(-\frac{c_s^2\rho_0 V_{cr}^2}{k_B T}n_b\right) = \Gamma(W_{cr,0})\exp\left(-\frac{c_s^2\rho_0 V_{cr}^2}{k_B T}n_b\right) , \tag{24}$$

where $\Gamma(W_{cr,0}) = \frac{3}{4\pi R_{cr}^3}\frac{k_B T}{2\pi\hbar}\exp\left(-\frac{W_{cr,0}}{k_B T}\right)$ .

If we recall that the connection $R_{cr}$ and $W_{cr,0}$ with $\delta_b$ and $P_{-,0}$ are given by (12) and (14),
The equation (24) has a simple solution at fixed values $\delta_b$ and $P_-$:

$$n_{b,satur} = \frac{\ln\left(\Gamma(W_{cr,0})\frac{c_s^2\rho_0 V_{cr}^2}{k_B T}t + 1\right)}{\frac{c_s^2\rho_0 V_{cr}^2}{k_B T}} . \tag{25}$$

Figure 6 shows the dependence $n_b(t)$ with and without the saturation effect. For a small period of time when:

$$\Gamma(W_{cr,0})\frac{c_s^2\rho_0 V_{cr}^2}{k_B T}t < 1, \tag{26}$$

the nanopore density increases linearly with time. Then, taking into account the saturation, the dependence $n_b(t)$ becomes logarithmic.

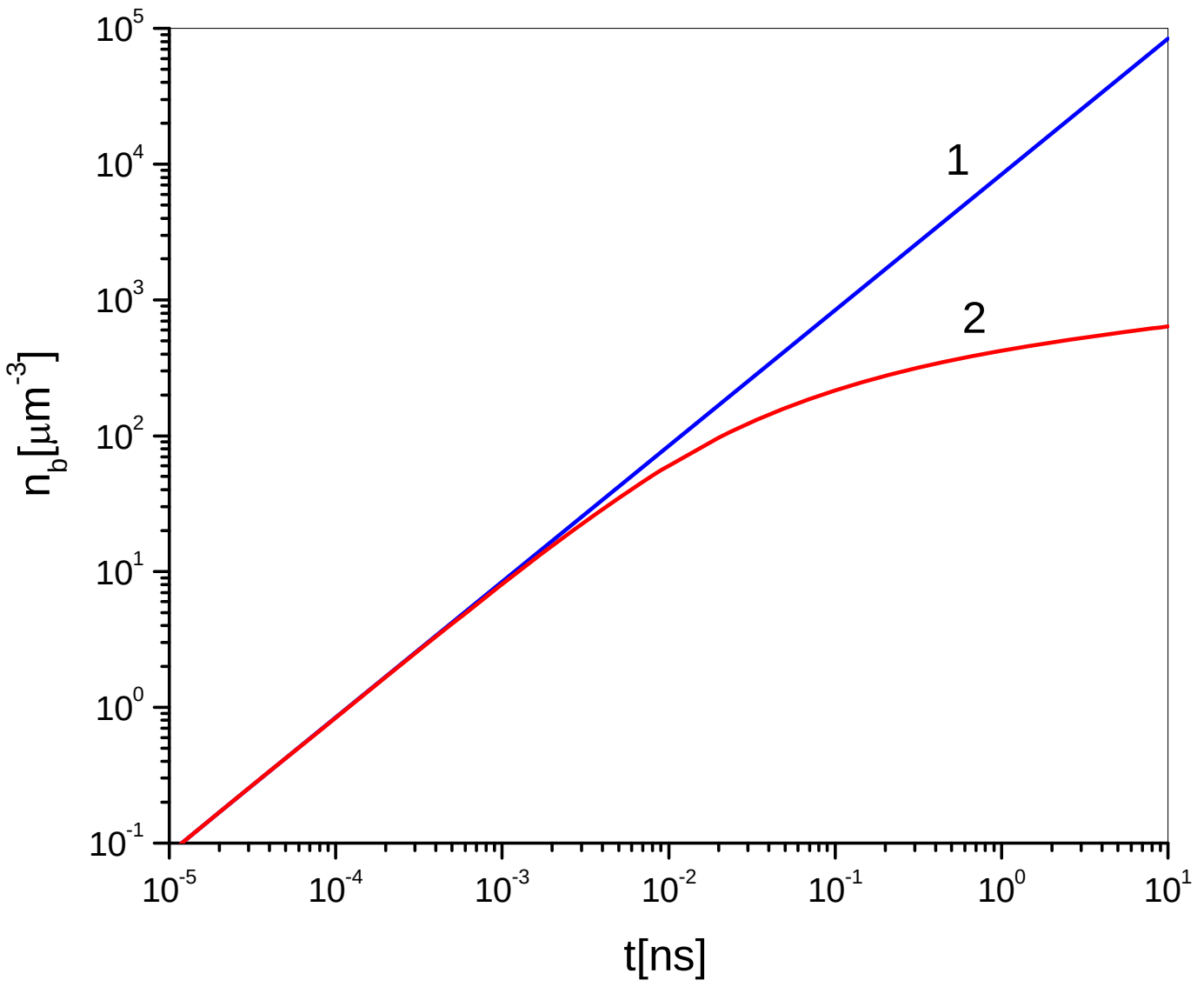

**Fig. 6.** The dependence of the emerging nanopores density on time. Line 1 - without considering the effect of saturation (formula (17)), Line 2 – with saturation (formula (25)). $\delta_b$ = 2.255nm ($R_{cr}$ = 3.2nm), $|P_{-,0}| = 30\text{MPa}$ .

Figure 7 shows the dependence of $n_b(t)$ at various values of $\delta_b$, when the saturation effect is taken into account. The changing of $\delta_b$ by 0.015nm (0.7%) causes the changing of the nanopores density by an order, while the size of the nanopores, remains almost the same, $R_{cr} \approx 3.2$ nm. Such substantial threshold dependence of the density of nanopores $n_b$ on $\delta_b$ allows to determine the parameters $\delta_b$ and $R_{cr}$ for the known value of the critical negative pressure at which cavitation begins. An optical method for detection of the nanopores appearance at the initial stages of cavitation development will be considered in the next section of the paper.

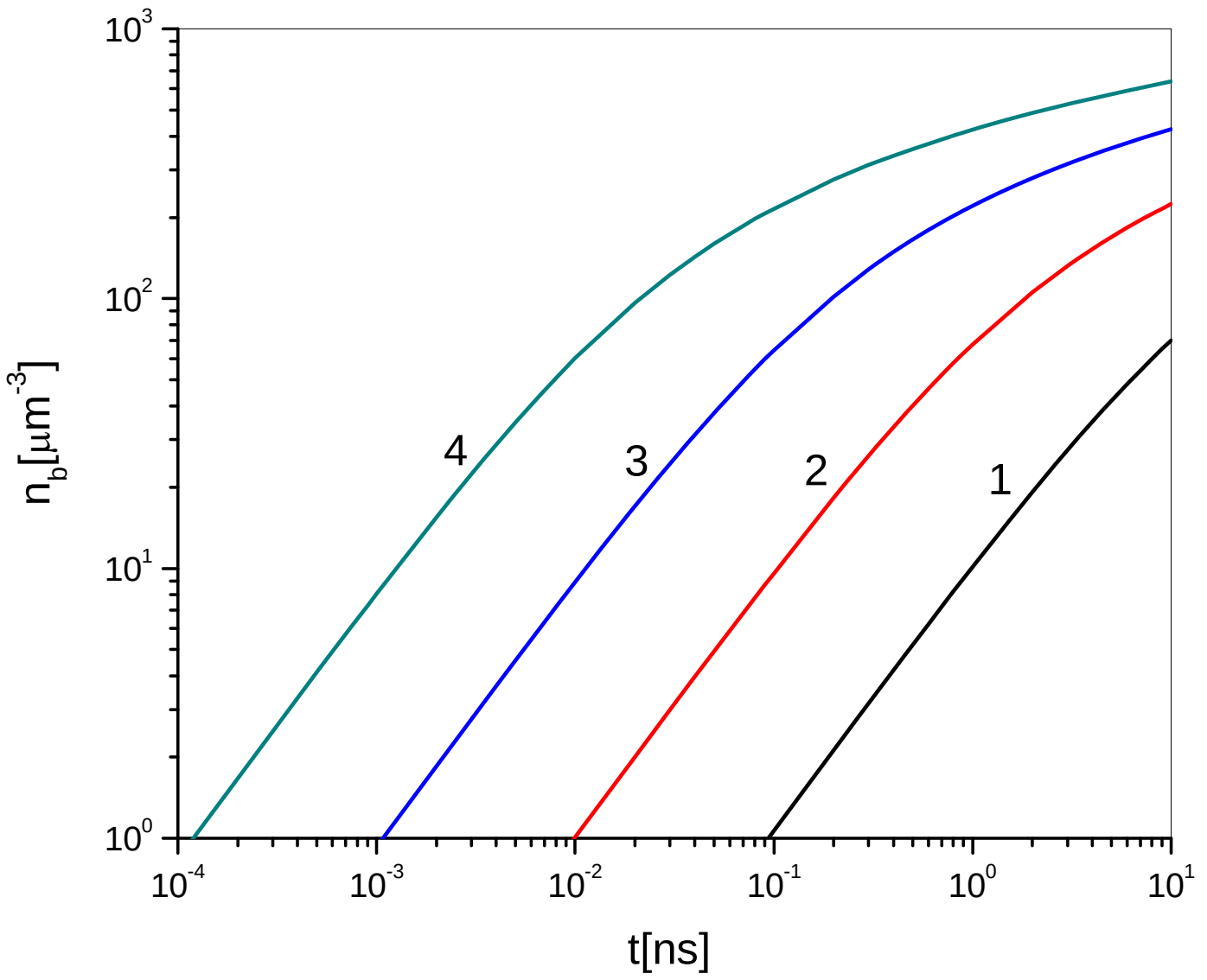

Fig. 7. The dpendence of the density of cavitation nanopores on time. Line 1 corresponds to $\delta_b$ = 2.24nm ( $R_{cr}$ =3.262nm), 2 – $\delta_b$ = 2.245nm ( $R_{cr}$ =3.249nm), 3 – $\delta_b$ =2.250nm ( $R_{cr}$ =3.235nm), 4 – $\delta_b$ = 2.255nm ( $R_{cr}$ =3.222nm). $\left|P_{-,0}\right| = 30\text{MPa}$ .

The estimate (25) holds if $\Gamma(W_{cr,0})\tau >> n_{b,satur}$ .

It should be noted that the saturation effect of the nanopores density occurs at any assumed dependence of the surface tension $\sigma_s(R_{cr})$.

## 3. Rayleigh scattering on the cavitation region emerging in liquids

One of the main problems of studying cavitation in different environments is the experimental determination of the critical value of the negative pressure at which it starts to form, and the frequency of microvoid formation in the liquid.

Conventional optical methods such as shadowgraph, schlieren, speckle photography, which are widely used in the hydrodynamic flows studies [19-23], do not allow for the detection of submicron nanopores in liquids. This, along with the substantial heterogeneity of the pressure field in the cavitation experiments, causes a wide range in the experimental values of the critical pressure, at which cavitation occurs.

It was shown in [24] that the Rayleigh scattering allows to observe the very beginning stages of cavitation development, when the emerging nanopores have not yet reached the size sufficient for detection by standard optical methods listed above.

In this section, we will show that the Rayleigh scattering off nanopores, emerging from the negative pressure regions of the liquid, can be used to detect cavities earlier in their development than through other optical methods.

The Rayleigh scattering by inhomogeneities in a medium occurs when their size is much smaller than the wavelength of light $\lambda$ [25]. Such inhomogeneities may be any fluctuations of density in a medium, including micropores. This means that for nanopores (cavitation nano-voids), the Rayleigh scattering is possible when the pore size satisfies $R_b << \lambda / n$, where $n$ is the refractive index of the medium, and $\lambda$ is the wavelength of light in vacuum. In this case, we can assume that a nanopore is located in the uniform electric field $E = E_0 \cdot e^{i\omega t}$, where $\omega = 2\pi c / \lambda$, and $c$ is the speed of light in vacuum. In the frequency range of visible light, $n \approx 1.33$ for water. In such a field, a spherical cavity of radius $a$ behaves like an oscillating dipole with the dipole moment

$$\vec{p}_b = \alpha \vec{E}_0 \cdot e^{i\omega t} \tag{27}$$

due to the periodic polarization of liquid on its borders. Here

$$\alpha = 4\pi\varepsilon_0 a^3 \frac{1 - n^2}{1 + 2n^2} \tag{28}$$

is the effective polarizability of the cavity in dielectric media [26], in which $\varepsilon \approx n^2$ is assumed for visible light in water.

To describe Rayleigh scattering in media, it is convenient to use the so-called scattering factor [26]:

$$\Upsilon_b(r,\theta) = \frac{I(r,\theta)}{I_I}, \tag{29}$$

where $I_I = \varepsilon_0 n c E_0^2 / 2$ is the intensity of the incident radiation, in which $E_0$ is the electric field amplitude, $I(r,\theta) = \frac{\omega^4 \alpha^2 \sin^2\theta}{16\pi^2 r^2 \varepsilon_0^2 c^4} I_I$ is the intensity of the scattered radiation such that it makes the angle $\theta$ with respect to the induced dipole vector, and $r$ is the distance from the nanopore to the observation point [27,28] (Fig. 7.). The corresponding scattering factors in the direction determined by angle $\theta$ and integrating over the solid angle are, correspondingly

$$\Upsilon_b(r,\theta) = \frac{I(r,\theta)}{I_I} = \frac{\omega^4 \alpha^2 \sin^2\theta}{16\pi^2 r^2 \varepsilon_0^2} = \frac{\pi^2 \alpha^2 \sin^2\theta}{\lambda^4 \varepsilon_0^2 r^2} \tag{30}$$

$$\Upsilon_{\Omega,b}(r) = \int \Upsilon_b(r,\theta) d\Omega = \frac{8\pi^3 \alpha^2}{3 r^2 \varepsilon_0^2 \lambda^4}. \tag{31}$$

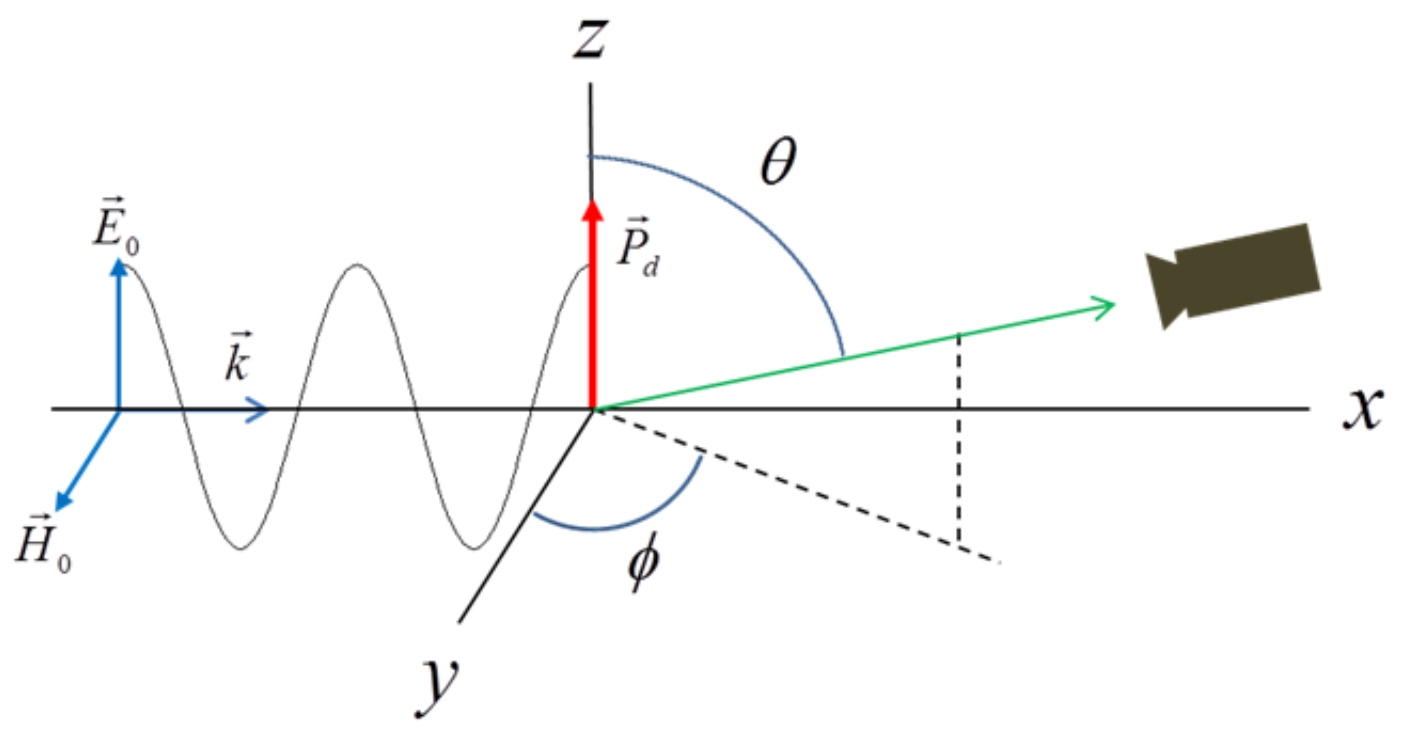

**Fig. 7. The** Scheme of Rayleigh scattering. $\vec{P}_d$ is the induced dipole moment.

If the cavitation nanopores are distributed randomly, and the average distance between them is greater than the wavelength $l_b > \lambda / n$, the scattered radiation is uncorrelated, and the scattering factor is proportional to the number of pores $N_b$ in the irradiated scattering volume:

$$\Upsilon_{N,b}(r,\theta) = N_b \cdot \Upsilon_b(r,\theta) \tag{32}$$

$$\Upsilon_{N,\Omega,b}(r) = N_b \cdot \Upsilon_{\Omega,b}(r) \quad . \tag{33}$$

The scattering off of cavitation micropores will be noticeable if it reaches or exceeds the level of Rayleigh scattering off of the background liquid. For the scattering by the background liquid molecules, when the characteristic size of the scattering region $L_s >> \lambda / n$, the complete mutual interference quenching of radiation scattered by individual molecules holds, and Rayleigh scattering is determined by thermal fluctuations [29,30]. Note that in our recent paper [24], we erroneously compared the scattering of a plane-polarized electromagnetic wave off nanopores to the scattering of natural polarized light off the background fluctuations of the water. Below, we will correct this error.

The Rayleigh scattering factor for a small volume of liquid $V$ , irradiated with a monochromatic plane electromagnetic wave, can be written in terms of $\theta$, the angle between the scattered and the incident radiations (Fig. 7), and also in terms of scattering over the full solid angle:

$$\Upsilon_{fluid}(r,\theta) = \frac{\pi^2 V}{4\lambda^4 r^2}\left(\rho \frac{\partial \varepsilon}{\partial \rho}\right)_T^2 \beta_T k_B T \sin^2\theta , \tag{34}$$

$$\Upsilon_{\Omega,fluid}(r) = \int \Upsilon_{fluid}(\theta,\phi,r) d\Omega = \frac{2\pi^3 V}{3\lambda^4 r^2}\left(\rho \frac{\partial \varepsilon}{\partial \rho}\right)_T^2 \beta_T k_B T \ . \tag{35}$$

Here, $\beta_T = -\left(\frac{1}{V}\frac{\partial V}{\partial p}\right)_T = 4.8 \cdot 10^{-10}\left[\frac{\mathrm{m}^2}{\mathrm{N}}\right]$ is the coefficient of volume expansion of water, $k_B$ is the Boltzmann constant, and $\left(\rho \frac{\partial \varepsilon}{\partial \rho}\right)_T \approx 1$.

The ratio of the Rayleigh scattering intensities off of cavitation micropores in volume $V$ and off of water of the same volume is independent of scattering angles and follows from (32)-(35):

$$\xi = \frac{\Upsilon_{N,b}(r,\theta)}{\Upsilon_{fluid}(r,\theta)} = \frac{\Upsilon_{N,\Omega,b}}{\Upsilon_{\Omega,fluid}} = \frac{64\pi^2}{\left(\rho \frac{\partial \varepsilon}{\partial \rho}\right)_T^2 \beta_T k_B T}\left(\frac{1-n^2}{1+2n^2}\right)^2 \cdot n_b a^6 \approx 9 \cdot 10^{-6} n_b a^6 \tag{36}$$

Here, $n_b = N_b / V$ is the density of the nanopores in $\mu m^{-3}$, $a$ is the pore radius in nanometers, and the numerical coefficients correspond to the water temperature of 300 K. Since $l_b > \lambda / n$ holds for the Rayleigh scattering off of cavitation bubbles, considering that $n_b \sim (\lambda / n)^{-3}$, we get $\xi \approx 8 \cdot 10^{-5} a^6$ for green light ($\lambda = 532$ nm). That is, when the size of the nanopores is $a > 4.8\,\text{nm}$ in volume illuminated by a laser, the Rayleigh scattering off of the nanopores exceeds the scattering off of the thermal fluctuations of water in the same volume.

Let us obtain the estimations for the Rayleigh scattering off nanopores in the case of the assumed surface tension dependence on the pore radius (9).

Without loss of generality, assume that the absolute value of the negative pressure increases linearly with time and then remains constant.

$$|P_-| = \begin{cases} P_0 \dfrac{t}{t_0} & 0 < t \le t_0 \\ P_0 & \text{t} > t_0 \end{cases} \tag{37}$$

Figure 8 shows examples of the time dependencies of the rate of generation of cavitation voids, the number of pores, and parameter $\xi$ for the following assumed parameters of the problem: $P_0 = 30\text{MPa}$, $\delta_b = 2.243\text{nm}$ (corresponding $R_{cr} = 3.25$ nm), $t_0 = 3\text{ns}$, and the surface tension coefficient $\sigma_{0s} = 0.072$ N/m .

Note that for the time of the order of tens of nanoseconds, pores can grow into much larger sizes (the rate of the expansion of nanopores is about 100-300 m/s [31,32]), but we disregard this fact for simplicity. If, in the process of growth, the size of the pores reaches the order of the laser wavelength, then the scattering ceases to be isotropic Rayleigh and becomes anisotropic Mie scattering (e.g. [26, 27]), which we do not consider.

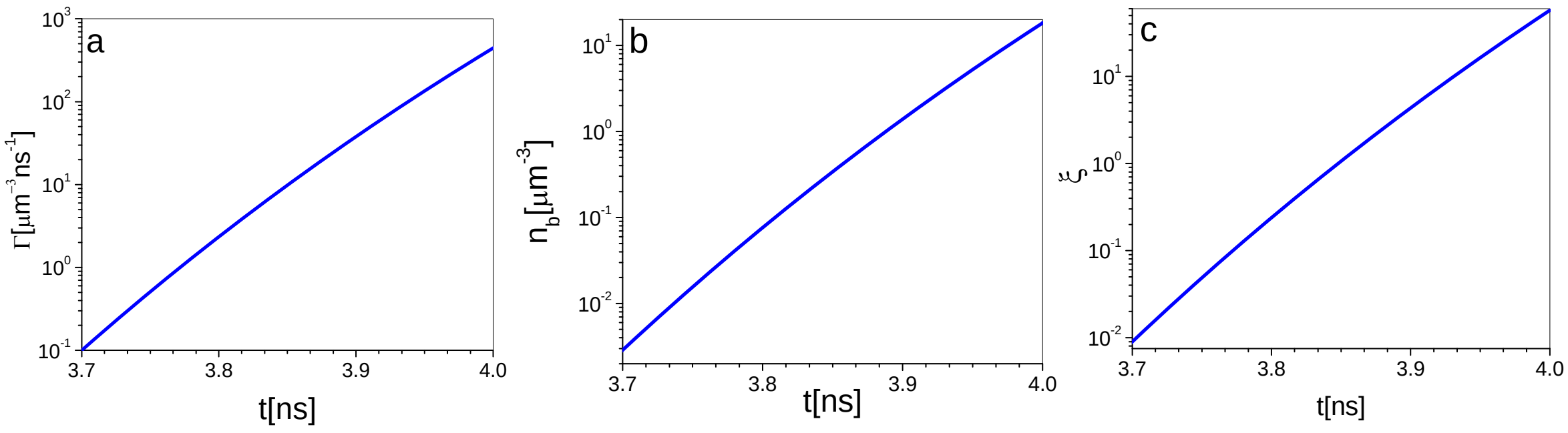

**Fig. 8**. Time dependencies of the generation rate of cavitation voids, the number of pores, and the parameter $\xi$ at $P_0 = 30\text{MPa}$, $\delta_b = 2.243\text{nm}$ (corresponding $R_{cr} = 3.25\,\text{nm}$), $t_0 = 3\text{ns}$.

Measuring of the critical negative pressure at which cavitation occurs and the rate of generation of nanopores allows to define the parameters $R_{cr}$ and $\delta_b$ that determine the theory described above.

## Conclusions

1. A theory of cavitation inception within the Zel'dovich-Fisher nucleation theory with taking into account the saturation of nanovoids generation is developed.

2. It is shown that the Rayleigh scattering allows the detection of cavitation in the early stages of its development.